# COOLING OF PARTICLE BEAMS IN STORAGE RINGS

*E.G.Bessonov[1]*

[1]*Lebedev Physical Institute RAS, Moscow, Russia*
*E-mail:bessonov@x4u.lebedev.ru*

PACS: 29.20.Dh, 07.85.Fv, 29.27.Eg.

## 1. INTRODUCTION

Beam cooling is the reduction of its normalized six dimensional (6D) emittance (hypervolume). The 6D emittance and the sums of 4D and 2D beam emittances

$$\in_{n,6D} = \iiint..\int dV_6 =$$
$$\iiint..\int dx \cdot dy \cdot dz \cdot dp_x \cdot dp_y \cdot dp_z, \quad (1)$$
$$\sum_i \in_{n,2D,i} = \sum_i \iint dx_i dp_{x_i},$$
$$\sum_{i \neq k} \in_{n,4D,i,k} = \sum_{i \neq k} \iiint\int dx_i dp_{x_i} dx_k dp_{x_k},$$

according to Liouville's theorem [1], are invariants for conservative Hamiltonian system. Here $dx_i|_{i=1,2,3} = dx$, $dy, dz$, the normalized and unnormalized emittances are connected by the relations: $\in_{n,x} = \beta\gamma \in_x$, $\in_x = \iint dx \cdot dx/dz$, $\beta = v/c$, $\gamma = 1/\sqrt{1-\beta^2}$. This is the consequence of the lack of the divergence for such systems in the 6D velocity $\vec{v}(dx/dt,...,dp_x/dt,...)$ ($div\,\vec{v} = 0$ and hence the derivative of the 6D density $d\rho/dt = 0$ for such system).

Note, that the term cooling does not refer to the temperature in thermodynamics. If the movement of particles in different directions is independent, then $\in_{n,6D} = \in_{n,x} \in_{n,z} \in_{n,s}$.

Emittance determines energy and angular spreads of the particles in the beam, its dimensions. Emittance and the number of particles in the beam determine the principal parameter of the beam, its normalized brightness $B_n = N/\in_{n,6D}$. If the density of the number of particles in the phase space is not homogeneous one then the idea of the density of the brightness can be introduced $b_n = dN/d\in_{n,6D}$. Multicycle injection and production of very dense particle beams in storage rings is possible only under conditions of 6D cooling.

The Liouville's theorem was proofed for arbitrary external fields (linear, nonlinear) and for:
1) conservative systems,
2) the intrabeam interactions are neglected,
3) the density is continuous function of particle position and momentum.

In reality the system has a friction, the intrabeam interactions occur, the number of particles and distance between particles are finite (countable set of particles).

## 2. METHODS OF THE BEAM COOLING

To cool a particle beam we must violate the conditions for which the Liouvilles theorem was proved. For this purpose we can introduce in the system:

1) A friction (non-conservatism) to convert the system into non-Hamiltonian one. Friction, in case of linear systems leads to appearance of damping decrements. There is a correlation between the decrements determined by Robinson's damping criterion [2], [3], [4]. It limits the rate of particle cooling in storage rings:

$$\alpha_x = -\frac{1}{2}\left[\frac{\overline{P}}{\varepsilon_s} + \frac{\partial \overline{P}}{\partial \varepsilon}\big|_s - \frac{d\overline{P}}{d\varepsilon}\big|_s\right], \quad \alpha_y = -\frac{1}{2}\frac{\overline{P}}{\varepsilon_s},$$
$$\alpha_s = -\frac{1}{2}\frac{d\overline{P}}{d\varepsilon}\big|_s, \quad \sum_i^3 \alpha_i = -\frac{\overline{P}}{\varepsilon_s} - \frac{1}{2}\frac{\partial \overline{P}}{\partial \varepsilon}\big|_s, \quad (2)$$

where $\overline{P} = \frac{1}{T_s}\oint P(x,t)dt$, $s$ relates to equilibrium value.

Coefficients $\alpha_i$ and their sum can be both negative and positive. Cooling in 6D phase space can be followed by damping in one direction and antidamping in another one. Fast 6D cooling occur if $\partial P/\partial\varepsilon|_s \gg \overline{P}/\varepsilon_s > 0$. The energy loss function must be increasing with energy for longitudinal cooling. Faster cooling can occur if the energy losses are nonlinear.

Note, that the expression (2) is valid if all $\alpha_i < \omega_i$, where $\omega_i$ is the mode frequency (oscillatory mode).

Decrements can be presented as $\alpha_i = -(\overline{P}/2c\varepsilon_s)\Im_i$. The sum of factors (partition numbers) $\Im_\Sigma = \sum_i \Im_i$ depends on the method of cooling. For radiation cooling $\Im_\Sigma = 4$, for stimulated radiation cooling $\Im_\Sigma \gg 1$. Damping time is $\tau_s \simeq \varepsilon_s/\overline{P}$ for ordinary cooling and $\tau_s \simeq \Delta\varepsilon_b/\overline{P} \ll \varepsilon_s/\overline{P}$ for stimulated radiation cooling, where $\Delta\varepsilon_b$ is the initial energy spread of the beam. The last expression is valid for selective methods of cooling based on interparticle spacing as well (see below).

Friction is necessary but not sufficient condition for damping. Damping of the beam in one plane and antidamping in another one can occur as well. The last case can be at the conditions of cooling.

2) Inelastic intrabeam interactions of particles in the beam can lead to cooling (with excitation of electronic or nuclear transitions).

3) A selectivity based on countable set of particles. In this case external forces act over small distances compared with the interparticle spacing both in ordinary 3D space or in 3D momentum space or in 6D phase space. Fast microwave electronics or thin and short laser-like wavelets (dimensions are less than interparticle distance) or monochromatic laser light for ion beams (using electronic or nuclear transitions) can be used.

At present the next cooling methods are known.

I. Cooling based on friction: 1) Radiation cooling,

2) Ionization cooling by energy losses in media (bremsstrahlung, ionization, excitation), 3) Electron cooling, 4) Cooling of relativistic ion beam by counter propagated monochromatic laser beam, 5) Radiative (Stimulated radiation) ion cooling by broadband laser beam.

II. Cooling based on interparticle spacing: 6) Stochastic cooling, 7) Optical stochastic cooling (OSC), 8) Transit-time OSC, 9) Enhanced optical cooling (EOC).

Below a brief description is given about each of mentioned methods of beam cooling in storage rings and linear accelerators.

*1) Radiation cooling.*

In 1946, D.Bohm, L.Foldy pointed out that the synchrotron radiation (SR) emitted by an electron in bending magnets of race-track synchrotrons with field index "n" leads to damping (0<n<3/4) or antidamping (n>3/4) of synchrotron oscillations [5]. In 1956 A.Kolomenskiy and A.Lebedev pointed out on the damping of the particle beam in the transverse plane [6], [7].

Quantum fluctuations in the energy loss were pointed out by A.Sokolov, I.Ternov in 1953 [8]. They were developed by M.Sands, A.Kolomenskiy, A.Lebedev in 1955, 1956 [9], [10]. Damping effect was quantified and made specific by K.Robinson in 1956 (Robinson damping criterion). Radiation cooling takes place in the electric and electromagnetic (laser) fields as well.

*2) Ionization cooling.*

In 1956, O'Neil G. pointed out on ionization cooling [11]. This idea was developed by other authors [3], [12], [13]. G.Budker in 1970 suggested to apply this idea to muon cooling [14]. According to Robinson damping criterion cooling and heating can occur because of the specific low of the energy loss in this case [15] - [17]. For muons it is rapidly decreasing with the energy for $\varepsilon < 300$ MeV but is slightly increasing for $\varepsilon > 300$ MeV (unfortunately the misleading term longitudinal cooling instead of damping was used in [17] at al.).

*3) Cooling of relativistic ion beam by counter propagated monochromatic laser beam ( laser cooling).*

The energy loss function of relativistic ions in the field of counter-propagated laser beam is very rapidly increasing one in the limits of the left part of narrow region of resonance interaction. It leads to fast damping in the longitudinal plane in this region. In general the friction forth in this case is not linear function of the ion energy and leads to non-exponential damping. Sweeping of the laser frequency is used for continuous interaction of ions with the laser beam until all ions decrease their energy to the synchronous one. Ordinary damping time in the transverse plane is negligible here. Beam emittance exchange can be used for fast transverse damping.

Here we use the idea of the energy loss function independently on the kind of the external electromagnetic field. In the case of the electromagnetic wave (dynamic undulator) and the ordinary magnetic undulator some peculiarity appear. If the wave co-propagate with the particle beam, then the wave accelerate the particles. We omit this case here.

In the frame of quantum theory the relativistic ion catch counter-propagating optical photon and decrease a little its momentum and velocity in the laboratory frame. At the same time the ion increases its mass on the value $\sim 2\gamma^2 \hbar \omega_l / c^2$. After emission the ion velocity is not changed in average but the energy and momentum are decreased essentially. Re-emission leads to heating (quantum excitation of oscillations).

First idea of laser cooling of ions in gaps was suggested in 1975 [18], [19]. Cooling in storage rings was developed by many authors both theoretically and experimentally [20]-[28].

*4) Electron cooling.*

The idea (1966) is in slowing down of swift particles through friction in matter moving with the average velocity of particles [29], [30]. It is similar to ionization cooling by co propagated cold media. It was realized for cooling of hot ion and antiproton beams by co propagated cold electron beams (Budker's cooling). The scheme is effective for cooling of nonrelativistic heavy particle beams.

*5) Radiative (stimulated radiation) cooling.*

Radiation cooling considered above is determined by Compton scattering cross-section of particles in the external fields. It is equal to $\sigma_T \simeq 6.65 \cdot 10^{-25}$ for electrons. At the same time the Rayleigh scattering cross-section of photons by not fully stripped ions (electronic transitions) or by fully stripped ions (nuclear transitions) is 10-15 orders higher. In this case the laser beam spectral bandwidth must correspond to the energy spread of the beam $\Delta\omega/\omega \simeq \Delta\varepsilon/\varepsilon_s$ (cooling by broadband laser beam). If the spectral distribution is linear decreasing function of frequency then Robinson damping criterion is valid and fast cooling will take place: $\Im_\Sigma = \Sigma_i \Im_i \gg 1$. Note, that linear distribution is not optimal one. Homogeneous distribution in the limits $\Delta\omega$ is more effective if maximum frequency corresponds to the synchronous ion energy. In this case ions will be gathered at minimum energy (corresponding to maximum laser frequency) [31]-[34].

*6) Stochastic cooling.*

The idea is due to Simon van der Meer (1972) [35], [36]. Pick-up takes signals from particles of the beam, which are amplified and go across the span of a storage ring so as to put an appropriate signal upon a kicker just as the particle arrives on its circular route to decrease the amplitude of the particle oscillation. Typical frequencies employed in stochastic cooling are ~5 GHz (~5cm). The distance between pick-up and kicker is $l = (2n+1)\lambda_b/4$, where $\lambda_b$ is the wavelength of betatron oscillations. Isochronous bend between pick-up and kicker must be used.

*7) Optical stochastic cooling (OSC).*

The idea is due to A.Mikhailichenko, M.Zolotorev (1993) [37]. The scheme is similar to ordinary stochastic one. Pick-up quadrupole undulator (PU) and ordinary kicker undulator (KU) are used. The undulator radiation (UR) wavelet (URW) emitted in the PU is amplified, go across the span of a storage ring so as to put an appropriate signal upon a KU just as the particle arrives on its circular route to decrease its energy. The amplitude of the magnetic field of the PU is increased with the deviation from the synchronous orbit. That is why the amplitude of the URW and the wavelength of the UR are increased as well. The change of the wavelength leads to decrease of the particle energy region of the resonance interaction with URW in the KU if the maximal deflection parameter of PU is ~1. Damping of betatron oscillations takes place as well if phase advance between undulators is $\psi_x^{bet} = 2\pi(k+1/2)$, $k = 1, 2, 3...$.

Such regime can be used to restrain particle beam from heating (the walls in the hypervolume). If the amplitude of the magnetic field of the KU is increased with the positive deviation from the synchronous orbit like in PU and stay near zero with negative one (nonlinear dependence), then the energy region of the resonance interaction will be increased.

*8) Transit-time method of OSC.*

The idea is due to M.S.Zolotorev, M.Zholentz (1994) [38]. Two identical ordinary undulators are used in their scheme. Lattice with non-zero local slippage factor between undulators is used. The phase of the particle in the URW is chosen zero for synchronous particle. The phase and the rate of the energy loss of particles are increased with the deviation of the particle energy from synchronous one. In this case particles loose energy by exponential low and the beam is cooled in the longitudinal direction. Damping in the transverse plain takes place if phase advance between undulators is chosen correct way.

*9) Enhansed Optical Cooling (EOC).*

Two identical ordinary undulators are used in this scheme. Lattice with local slippage factor $\eta_{c,l} = 0$ and phase advance between undulators $\psi_x^{bet} \cong 2\pi(k+1/2)$, $k = 1, 2, 3...$ must be used [39]-[43]. Special optical system including lenses, optical line with variable time delay, optical filter and movable screen located in the image plane are used. URW's emitted in the PU are focused by a lens on the image plane. Movable screen overlap the image of the beam in the initial state. Then it open URWs emitted by particles with maximal deviations of their position from synchronous one, go to the image of the synchronous particle and stopped (see Fig.1). The phase of the particle in the field of URW in KU is chosen so that the rate of the energy loss for particles is maximum. It does not depend on the deviation of the particle energy from synchronous one for open particles and equal zero for screened ones (fast nonexponential cooling). In this case the beam is cooled in longitudinal and transverse directions.

In the EOC scheme every particle enters the kicker undulator together with the head of the emitted by the particle URW in decelerating phase and looses its energy. URWs emitted by particles in the pickup undulator with positive deviations from the synchronous orbit are selected by movable screen in the image plane of the optical system. The change of the amplitude of the particle betatron oscillations: $\delta A_x^2 = -2x_{\beta,k}\delta x_\eta + (\delta x_\eta)^2 \simeq -2x_{\beta,k}\delta x_\eta < 0$ at $x_{\beta,k} < 0$, $\delta x_\eta < 0$, where $x_{\beta,k}$ is the initial particle deviation from it's closed orbit in kicker undulator; $\delta x_\eta = \eta_x \beta^{-2}(\delta E/E)$ is the change of it's closed orbit position; $\eta_x$ is the dispersion function in the storage ring; $\beta$ is the normalized velocity. Using short and narrow laser-like wavelets $\sim M\lambda_{1,\min} \times 2\sigma_w$ per-mits to give a chance to control the behavior of individual particles in the storage rings without significant disturbing another ones, where $M$ is the number of the undulator periods, $\sigma_w = \sqrt{Z_R \lambda_{1,\min}/4\pi}$, $2Z_R = M\lambda_u = L_u$, $Z_R$ is the Rayleigh length, $\lambda_{1,\min}$ is the minimum UR wavelength.

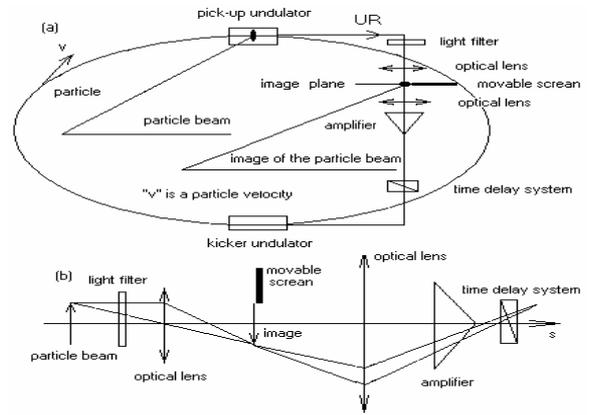

*Fig.1. The scheme of the EOC of particle beams in a storage rings (a) and unwrapped optical scheme (b).*

Small local slippage factor between undulators can be used as well [43].

The angular resolution of an electron bunch by an optical system is $\delta\varphi_{res} \cong 1.22\lambda_{1,\min}/D$, where $D$ is a minimum value between the diameter of the first lens and the diameter of the URW $\sigma_{w,b} \simeq (\Delta\theta)_c l$ at the position of the lens, $l$ is the distance between pickup undulator and the lens, $(\Delta\theta)_c = (1/\gamma)\sqrt{(1+K^2)/M}$ the angular range of the being used URW. The space resolution of the optical system is $\delta x_{res} \cong \delta\varphi_{res} l$ or

$$\delta x_{res} \cong 1.22\lambda_{1,\min}/(\Delta\theta)_c = 0.86\sqrt{\lambda_{1,\min}L_u}. \quad (3)$$

*10. Cooling in linear accelerators.*

Electron beams can be cooled in linear accelerators if external fields (undulators, electromagnetic waves) producing radiation friction forces will be distributed along the axes of these accelerators. The physics of cooling of electron beams under conditions of linear acceleration is similar to circular one. The electron looses its momentum in the direction of the particle velocity and therefore includes transverse and longitudinal momentum losses. The reacceleration of the electron beams in accelerating structures of linear accelerators restores the longitudinal momentum. The transverse emittance is reduced by 1/e with as little as 2ε of the total energy exchange. Cooling in the longitudinal plane is if the energy losses are increased with the particle energy.

First, the effect of undulator/wiggler radiation damping on the transverse beam emittance was studied by A.Ting and P.Sprangle for linear accelerators based on inverse free-electron lasers [44]. In [45], the radio frequency linear accelerators were considered for this goal. In [46], [47] and in [48], the case of linear acceleration was investigated, where a laser beam was used instead of a wiggler. General formulas in this cases are similar. Some peculiarities are in the hardness of the emitted radiation which determines the energy spread and transverse emittance of being cooled beams. The hardness of the backward scattered laser radiation is more high then wiggler radiation. That is why laser beams for damping can be used at small (~1 GeV) electron energies. Based on selective energy losses of ions in the laser beam and high cross-section (Rayleigh scattering) an effective cooling of ion beams can be used in linear accelerators through fast damping in the longitudinal plane.

## 3. POSSIBLE APPLICATIONS OF BEING COOLED BEAMS

1. Colliding beams ($\bar{p}$, $\mu^{\pm}$, $e^{+}$, ion beams).

2. Light sources based on relativistic particle beams (Backward Rayleigh scattering sources and quantum generators based on being cooled ion beams; SR, UR and backward Compton scattering sources based on electron beams).

3. Inertial fusion.

4. Crystalline ion beams.

"The cooling rings supply very dense beam either for direct use in precision experiments or for injection into big, high luminosity machines like the Tevatron or the projected next Linear Colliders and muon Colliders or the proposed Neutrino Factories" [49], [50], [51].

## 4. CONCLUSION

Much has been achieved with beam cooling, including:

1. All discoveries made with e+e- machines since 1960s (including charm, tau, synchrotron light physics) with radiation cooling.

2. W, Z, top, anti-H with stochastic cooling.

3. Beams of unprecedented brightness by electron cooling.

4. Special ion beams of still higher brightness by laser cooling.

5. A wide field of fascinating accelerator physics.

"In summary, we can say, beam cooling has led, leads, and will lead to spectacular results. In addition beam cooling was , is, and will be fun! What more do you want?" *(D.Mohl, A.M.Sessler, see [50]).*

Supported by RFBR under grant No 05-02-17162a

**COOLING OF PARTICLE BEAMS IN STORAGE RINGS**

*E.G.Bessonov*

Methods of particle beam cooling are reviewed.